# Catalyst design using actively learned machine with non-*ab initio* input features towards $CO_2$ reduction reactions

Juhwan Noh, Jaehoon Kim, Seoin Back, and Yousung Jung*

*Graduate School of EEWS, Korea Advanced Institute of Science and Technology (KAIST),*

*291 Daehakro, Daejeon 305-701, Korea*

*E-mail: ysjn@kaist.ac.kr

-Abstract-

In conventional chemisorption model, the d-band center theory (augmented sometimes with the upper edge of d-band for imporved accuarcy) plays a central role in predicting adsorption energies and catalytic activity as a function of d-band center of the solid surfaces, but it requires density functional calculations that can be quite costly for large scale screening purposes of materials. In this work, we propose to use the d-band width of the muffin-tin orbital theory (to account for local coordination environment) plus electronegativity (to account for adsorbate renormalization) as a simple set of alternative descriptors for chemisorption, which do not demand the ab initio calculations. This pair of descriptors are then combined with machine learning methods, namely, artificial neural network (ANN) and kernel ridge regression (KRR), to allow large scale materials screenings. We show, for a toy set of 263 alloy systems, that the CO adsorption energy can be predicted with a remarkably small mean absolute deviation error of 0.05 eV, a significantly improved result as compared to 0.13 eV obtained with descriptors including costly d-band center calculations in literature. We achieved this high accuracy by utilizing an active learning algorithm, without which the accuracy was 0.18 eV otherwise. As a practical application of this machine, we identified Cu3Y@Cu as a highly active and cost-effective electrochemical CO2 reduction catalyst to produce CO with the overpotential 0.37 V lower than Au catalyst.

# 1. Introduction

Understanding and predicting the energetics associated with bond-forming and bond-breaking reactions occurring on the surface of solid materials is the central theme in heterogeneous catalysis. Among many other catalysis theories, in particular, Sabatier principle[1] is an important simple concept that states a chemisorption strength of key reaction intermediates on catalyst surfaces should be just right to maximize the catalytic activity; either too weak or too strong binding leads to an insufficient activation of reactant or a great difficulty of product desorption after completion of the catalysis, respectively, and therefore a usual activity volcano relation can be plotted as a function of binding energies[2]. In a series of pioneering works, Nørskov and co-workers suggested a way to understand the chemisorption of reaction adsorbates in terms of the surface electronic structure of the materials in a so-called d-band theory [3]. Here, the essence is that the binding energy of an adsorbate to a metal surface is largely dependent on the electronic structure of the surface itself, namely, the d-band centre of the surface rather than the entire detailed density of states (DOS).

With the recent progress of electronic structure methods (mainly density functional theory calculations for solids) that can now give reliable electronic structure and binding energetics, the d-band center theory, along with the scaling relations that exist between the binding energies of related adsorbates, has been successfully applied to understanding and predicting new materials for many different applications[4-8]. However, exceptions were also found in which a usual d-band center trend cannot explain the activity measured [9-11]. The main cause of the latter exceptions was the lack of consideration on the spread in energy states, and for those cases, correlations between d-band center and the activity were improved by introducing the d-band width ($W_d$)[11] and upper edge of d-band ($E_u$)[10].

Recently, instead of energetic descriptors, an alternative metric to describe the activity of the catalyst based on the local geometric features of the active sites has been proposed, namely, the generalized coordination numbers[12, 13]. These approaches yielded simple coordination-activity plots that predicted the optimal geometric structure of platinum nanoparticles, which were then experimentally verified[14]. An additional advantage of this descriptor is that it does not require expensive *ab initio* calculations of projected density of states (pDOS) as in d-band theory, so that the method can be applied, in principle, to a massive screening of large database. Nonetheless, these generalized coordination numbers are not straightforward to extend to alloy systems, at least in its current form, since they cannot distinguish different electronic structures associated with different metal atoms in alloys. In this sense, it would be helpful to have a descriptor that can describe the local coordination environments as well as the electronic structure of constituting metal atoms when describing metal alloys. Indeed, an approach to satisfy the latter two aspects has been proposed in an orbitalwise coordination number[15] although it still requires time consuming *ab initio* calculated geometries to get high accuracy.

In this work, as a simple quantity to describe the local coordination and electronic structure modifications in alloys that does not require *ab initio* calculations, we propose to use the d-band width within the linear muffin-tin orbital (LMTO) theory[16]. Unlike the usual d-band width obtained from density functional calculations that is a bulk property of the slab, the LMTO d-band width in practice can capture the local electronic structure due to a truncation of the interatomic couplings up to the second nearest neighbor atoms. Using this LMTO-based d-band width, we construct, as a toy problem, a chemisorption model to compute the binding energy of *CO on various metal alloys. The idea is to establish a functional relation between the simple yet analytical LMTO d-band width and *CO binding energy, and perform large-scale screening using these non-*ab initio* descriptors to find a material with an optimal *CO binding for efficient $CO_2$ reduction reaction (CRR). Although there are many linear models between the descriptors such as d-band center (sometimes augmented by the upper edge of d-band) and the binding energy of adsorbate [10, 11, 17], to increase the prediction accuracy we here adopt the machine learning techniques to incorporate the potential nonlinear correlation between the descriptors and binding energies.

We note two machine learning models to predict *CO binding energy in literature using simple descriptors, with 13 electronic structure based descriptors in one case[18], and 2 local geometric features in the other case[19]. Incidentally, the authors, in both reports, obtained the similar mean absolute deviation error of 0.13 eV in predicting the *CO binding energy for various alloy systems despite the very different input features (13 electronic vs. 2 geometric), demonstrating the importance of proper feature selection for improved learning efficiency. It can also be noted that the input features in both machines still required *ab initio* calculations to relax geometries and/or to obtain accurate electronic structures of the materials. In a recent very interesting reaction network study[20], non-*ab initio* extended connectivity fingerprints (ECFP) based on Gaussian process (GP) model were used to predict the formation energies of ~90 surface intermediates species, with a final goal of identifying the most likely reaction pathways of syngas formation on rhodium (111), although a potential transferability limitation of ECFP for different adsorption sites were noted in which *ab initio* calculations would still be needed. Finding non-*ab initio* input features representing local environment is thus of significant current interests in heterogeneous catalysis.

In this work, we propose to use two non-*ab initio* input features, i.e., LMTO d-band width and electronegativity, as an easy-to-compute model to predict *CO binding energy on various alloy systems. Combining the latter descriptors and utilizing the latest active learning algorithms, we obtain root mean square error (RMSE) of 0.05 eV. As an application of the machine thus obtained to screen transition metal based alloy catalysts for CRR, we identified two promising catalysts, $Cu_3Zr@Cu$ and $Cu_3Y@Cu$, with higher activity than the most active Au based catalysts.

## 2. Methods

### 2.1 Non-*ab initio* descriptors for chemisorption

Selecting proper descriptor is one of the most important tasks in machine learning since it determines the learning efficiency as well as the prediction power. While most current descriptors for chemisorption models in machine learning demands *ab initio* calculations such as d-band center and its higher-order moments[19], our main focus is to utilize the non-*ab initio* based descriptors. As proposed by Nørskov and co-workers[3], the surface-adsorbate binding process can be decomposed into two effects; coupling of adsorbates with the sp- and d-bands of catalyst surface. For the former, based on an empirical correlation between the sp-coupling and surface-adsorbate bonding distance[17, 21-23], the sp-coupling term is usually estimated as a geometric mean of electronegativity for the first coordination shell ($\chi = \left[\prod_{i \in 1st\ n.n.}^{N} \chi_i\right]^{1/N}$), and we used the same practice. For the latter, instead of the conventional *ab initio* d-band center, we here propose to use non-*ab initio* analytical expression of d-band width within the LMTO theory, denoted as $W_d^{LTMO}$, (see Eq. S4 in ESI). While $W_d^{LTMO}$ does not contain the information about the center position of the d-band, there are two advantages of using it as a descriptor for large-scale screening purposes. First, it effectively captures the local chemical environment of the d-state for chemical events due to the use of interatomic coupling terms within the second nearest neighbor atoms from the active site, similar concepts used in the generalized coordination number[12, 13, 24]. Secondly, it does not require DFT calculations since the analytical expression for $W_d^{LTMO}$ can be evaluated based on the tabulated values for a given composition (See ESI).

### 2.2 Active learning

In machine learning, most of the computational cost in building the model usually occurs in generating the reference data in the training set and running the cost-function minimizations. Therefore, it is of significant practical interest to reduce the training set size as small as possible without compromising the representability of the system. Active learning is an algorithm in which the machine can point out samples with maximal information about the target function[25], and is widely used currently in classification/filtering, speech recognition, information extraction, computational biology, and etc, for example[26]. In this work, we utilize active learning in the catalyst design application to choose the minimal list of samples for training that can represent the given class of alloy materials considered.

We used two types of machine learning methods, neural network (NN) and kernel-ridge-regression (KRR) methods, as described in detail in the next Computational Details section. For active neural network learning[25], we used the ensemble NN model. In this method, one constructs multiple NN models in an ensemble (5 in our case) based on the same training set but optimized with different initializations, identifies examples in the test set characterized by the largest variance (or ambiguity) within the ensemble, and includes these new examples in the next training set for further learning. Since this algorithm does not require the label (*CO binding energy), we denote this method as *ensemble NN without label*. If one already has labels for the test set, improved accuracy might be expected by computing the actual residual, or error, (difference between the ensemble-averaged model-predicted values and true labels) and identifying an example with the largest residual. We denote it as *ensemble NN with label*.

For KRR, since there is an analytical unique solution for given training samples and hyper-parameters, methods similar to ensemble NN cannot be constructed. Instead, there are other flavours of active learning algorithms for KRR in literature[27-30], and in this work, we used the residual regression model. In this algorithm, one first obtains a *CO binding energy predictor with the training set (as in conventional KRR) with a certain training set error. Next, one constructs an error predictor with the same training set using the latter training set error as output data. This error predictor is then used to identify samples that are most different from the existing training set. In other words, one estimates the errors of all test samples using this error predictor, and find samples with the lowest absolute value of the generated outputs for further learning. Since this algorithm does not need labels of the test set, we denote this method as *active KRR without label*[29]. For a similar reason considered in *ensemble NN with label*, since in the present case all labels of the test set are available, we also constructed another active KRR model utilizing the labels of the test set, denoted as *active KRR with label*. Here, one includes samples with the highest absolute value of the residual of the error predictor on the test set into the next training set. The residual of the error predictor is defined as a difference between the output of the error predictor and the error calculated from the *CO binding energy predictor.

## 3. Models and Computational Details

### 3.1 Descriptor Evaluations

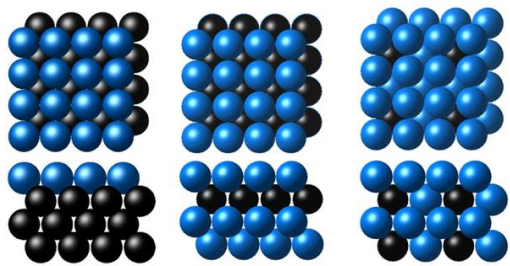

Fig. 1 Three surface overlayer alloy models considered in this study (from left to right): X@M, M-X@M, and M$_3$X@M (Blue and black balls denote M and X metals, respectively).

As a toy problem of predicting the *CO binding energy on the fcc(100) slabs, we considered surface overlayers in the form of X@M, M-X@M, and M$_3$X@M, where M=Ag, Au, Cu, Ni, Pd, Pt, and X is the 3d, 4d and 5d transition metals (total 263 samples), as shown in Fig. 1 and taken from [18, 19]. Calculated descriptors ($\chi$ and $W_d^{LMTO}$) for machine learning for the latter set are listed in Table S3 of ESI.

Counting of the first and second nearest neighboring atoms were defined layer by layer using the two topmost metal layers. In other words, for the first layer, around the binding site, there are 4-atoms in the nearest and 4-atoms in the second nearest neighbors on the basis of distance from the binding site. It can be similarly applied to the second layer; the number of the first and second nearest neighboring atoms around the binding site in the second layer would be 4 and 8. On this definition of coordination numbers, $\chi$ was calculated for both Mulliken ($\chi_M$) and Pauling scales ($\chi_P$). Details on the estimation of $W_d^{LTMO}$ are described in ESI, but we emphasize that $W_d^{LMTO}$ can be obtained without ab initio geometry relaxations, unlike in previous approaches[18] since the interatomic distances of alloy models are estimated using the Vegard's law[17] for the two topmost layers. More on the latter calculation details are shown in ESI.

Other quantities used for comparison and training/test such as $d_c$, $W_d^{cal}$ and *CO adsorption energy ($ECO_{cal}$) are taken from ref. [18]. The upper edge of d-band ($E_u$) is defined as $d_c + W_d/2$, and for clarity, $E_u^{LMTO}$ is defined as $d_c + W_d^{LMTO}/2$ and $E_u^{cal}$ is defined as $d_c + W_d^{cal}/2$.

**3.2 Machine learning models**

For all NN models, MATLAB R2015b Neural Network Toolbox ™ [31] was used with Parallel Computing Toolbox ™ [32]. For the training algorithm, Levenberg-Marquardt training algorithm [33-35], a kind of the backpropagation algorithms, was used with the hyperbolic-tangent activation function. For the conventional ANN, both single hidden layer (SHL ANN) and double hidden layer (DHL ANN) models were tested. The SHL ANN was trained with the number of nodes in hidden layer increasing from 4 to 20, and for the DHL ANN, a second hidden layer with 4 nodes was added to the SHL ANN. Total 263 data were randomly divided into three parts; training, validation and test sets with the ratio of 60:15:25 %.

To implement the ensemble NN, 5 independent DHL ANN models were constructed with 4 samples randomly chosen as an initial training set. During the active learning process, 2 additional samples were identified in each iteration from the untrained samples and added to the training set until the final training set reaches 60% of total data.

For all KRR models, the conventional KRR method (non-active KRR)[36] with the kernel function, $k(x_u, x_v) = \exp(-\|x_u - x_v\|_2/\sigma)$, was used. To reduce the computational cost of hyper-parameter optimizations, we explicitly fixed the kernel width as $\sigma$=0.5 and regularization factor to be 0.005. For active KRR models, the same procedure in ensemble NN was applied. For all machine learning models, 4800 independent trials were applied to remove randomness from initial training sets.

**3.3 DFT calculations**

We performed DFT calculations for selected candidates using Vienna Ab-initio Simulation Package (VASP)[37] with the projector-augmented wave (PAW)[38] and the revised Perdew-Burke-Ernzerhof (RPBE) exchange-correlational functional [39, 40]. The energy cut-off for plane-wave basis set was 500 eV, and k-points were sampled using the (8 × 4 × 1) Monkhorst-Pack mesh [41]. We modelled the fcc(100) slabs with (4 × 2) atom containing surface unit cell and 4 layers. 15Å vacuum was added to minimize the interaction between periodic images in z-direction. Topmost 2 layers were relaxed until the residual force on each atom becomes less than 0.05 eV/Å. Free energy of reaction intermediates on the surface was obtained by using harmonic oscillator approximation at 298.15K implemented in Atomic Simulation Environment (ASE) program [42], and free energy of gas molecule was obtained using ideal gas approximation at 298.15K implemented in ASE. To correct the systematic error of RPBE formation energies of relevant reactions compared to the experimental data, we added +0.45 eV correction for $CO_2$ molecule[43]. We also applied approximate solvation corrections for *CO (-0.10 eV) and *COOH (-0.25 eV) to account for the effect of solvation.

## 4. Results and Discussion

### 4.1. Local chemical environment description

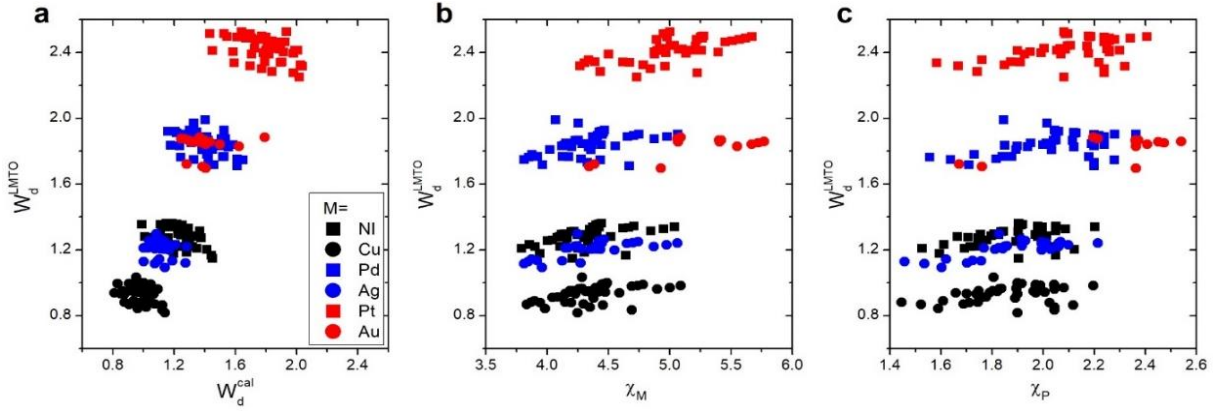

Fig. 2. (a) Comparison of LMTO-based d-band widths ($W_d^{LMTO}$) to the DFT calculated values ($W_d^{cal}$) for various alloy models, X@M, M-X@M, and $M_3$X@M (see Computational details section). The d-band widths are normalized to 1 for pure Cu for easy comparison. Data distributions of $W_d^{LMTO}$ versus two types of electronegativities, (b) Mulliken $\chi_M$ and (c) Pauling $\chi_P$.

We first compare the LMTO-based d-band widths, $W_d^{LMTO}$, to the DFT d-band widths, $W_d^{cal}$. Two main points can be drawn from Fig. 2a. First, $W_d^{LMTO}$ shows qualitatively the same trend as $W_d^{cal}$. Second, many different materials are clustered around the similar $W_d^{LMTO}$ or $W_d^{cal}$ values, but they are largely resolved by introducing both $\chi_M$ and $\chi_P$ as shown in Fig. 2b and 2c.

Using these two selected descriptors, the performances of various machine learning models are summarized in Fig. 3. Three points are noteworthy:

(1) Interestingly, the LMTO-based d-band width ($W_d^{LMTO}$) alone even performs quite well (RMSE = 0.07 eV). In addition, the LMTO-based d-band width consistently yields a lower RMSE than *ab initio*-based d-band width by 0.05–0.15 eV, when combined with $\chi$. This suggests that the local concept involved in the evaluation of $W_d^{LMTO}$ (interactions up to 2$^{nd}$ nearest neighboring atoms in the surface and subsurface layers) helps to correlate better with the binding affinity as compared to the bulk surface quantity ($W_d^{cal}$). The localized nature of $W_d^{LMTO}$ can also be confirmed in Fig. 4, in which the core@M alloys with the same M species are all clustered around the similar region, whereas the distribution of $W_d^{cal}$ is much broader for the same M. This clustering is a helpful feature for active learning since it becomes easier to choose a new data that is most different from the existing training set data.

(2) For all combinations of descriptors shown in Fig. 3a, KRR (0.05–0.37 eV) performs consistently better than ANN (0.21–0.45 eV) for the present chemistry.

(3) The actively learned KRR enhances the accuracy of the model significantly, lowering the RMSE by 0.13 eV compared to conventional KRR (from 0.18 to 0.05 eV) for the best case. More detailed discussions on the effects of active learning on the RMSE variances and accuracy for both ANN and KRR will be given later.

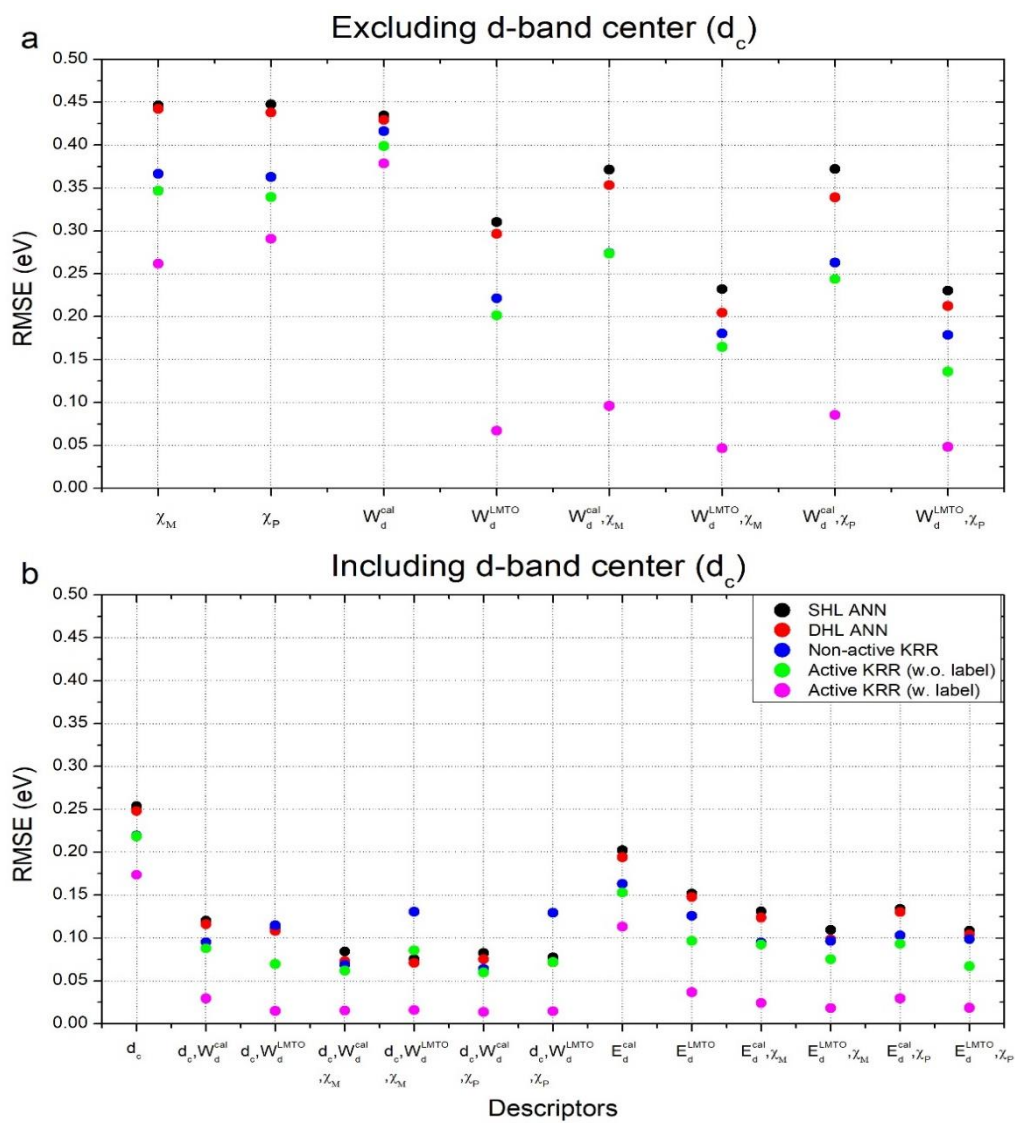

Fig. 3 Performance of various machine learning models with different descriptors (a) without *ab initio* d-band center, and (b) with d-band center. All RMSE values were calculated for the entire data set.

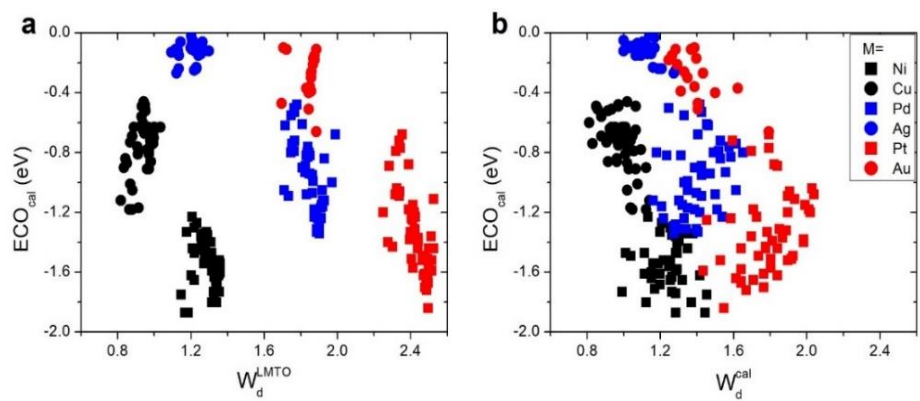

Fig. 4 Data distributions of *CO binding energies versus (a) $W_d^{LMTO}$ and (b) $W_d^{cal}$.

Combining all these results, therefore, the best chemisorption model without any *ab initio* inputs is the active KRR model based on the pair of ($W_d^{LMTO}$, $\chi_P$) descriptors with RMSE of 0.05 eV. This can be compared with the previous results (0.13 eV) using ANN with *ab initio* based parameters and geometries.

Since it is well known that the d-band center is the single best descriptor for the chemisorption on the catalyst surfaces, we also considered the models that include the conventional d-band center although it requires costly *ab initio* calculations (Fig. 3b-c). As expected, for all machine learning methods and combinations of descriptors, inclusion of d-band center improves the accuracy significantly, with the best model being the active KRR with any combinations of descriptors with RMSE ~ 0.02 eV. It is remarkable, however, to note that the difference between cost-effective LMTO d-band width model (0.05 eV) and expensive d-band center model is only 0.03 eV.

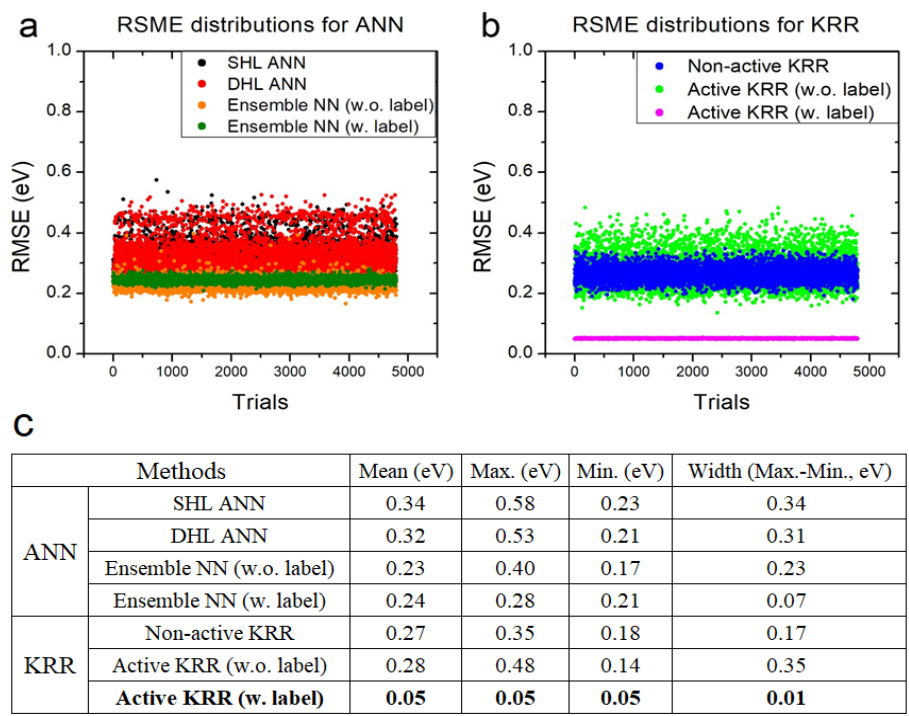

Fig. 5 RMSE distributions for (a) ANN and (b) KRR, and (c) their details for all trials. All statistics are for the machine learning model with the ($W_d^{LMTO}$, $\chi_P$) descriptors.

Next, we systematically analyzed the results with and without active learning techniques which are summarized in Fig. 5 with the raw RMSE data of 4800 independent trials for ANN and KRR. For ANN, it is clear that the widths of distributions are substantially reduced by applying active learning algorithm from 0.31 eV for DHL ANN to 0.07 eV for ensemble NN (w. labels). The similar behavior is also seen in KRR, in which the width of distribution decreased from 0.17 eV for non-active KRR to 0.01 eV for active learning with labels. Interestingly, in the absence of the labels, the width increases with active learning (albeit still with improved final accuracy). Although, as shown in Fig. 5c, the effects of active learning are much more pronounced for KRR (0.18 → 0.05 eV) than for ANN (0.21 → 0.17 eV), it is possible that the use of different active learning algorithm for ANN other than the ensemble method used here may further enhance the resulting accuracy.

As a practical application of the actively learned chemisorption model with $W_d^{LMTO}$ and $\chi_P$ as descriptors, we screened over 372 transition metal-based alloys (including 263 structures used in learning) with structures shown in Fig. 1 to find active CRR catalysts to produce CO. Particularly, it has been suggested based on DFT calculations that *CO binding energy is a key descriptor for the catalytic activity of $CO_2$ reduction[43], and the current density measurements on various metal catalysts indeed showed the volcano-shaped relation of activity with respect to *CO binding energy.[44] Currently, Au catalysts are reported to be the best single component catalyst for converting $CO_2$ into CO, but alternative cost-effective catalysts are in need due to the high cost and a scarcity of Au. To replace Au catalysts, one thus needs to develop catalysts with strong $E_{COOH}$ to facilitate the activation of $CO_2$, but not too strong $E_{CO}$ to easily remove the product. Considering that the optimal *CO binding energy to achieve facile *COOH formation and *CO desorption is approximately -0.5 eV based on the scaling relation and the volcano plot[43], we selected candidates of which *CO binding energies are in the range of -0.60 ~ -0.43 eV.[18, 43]

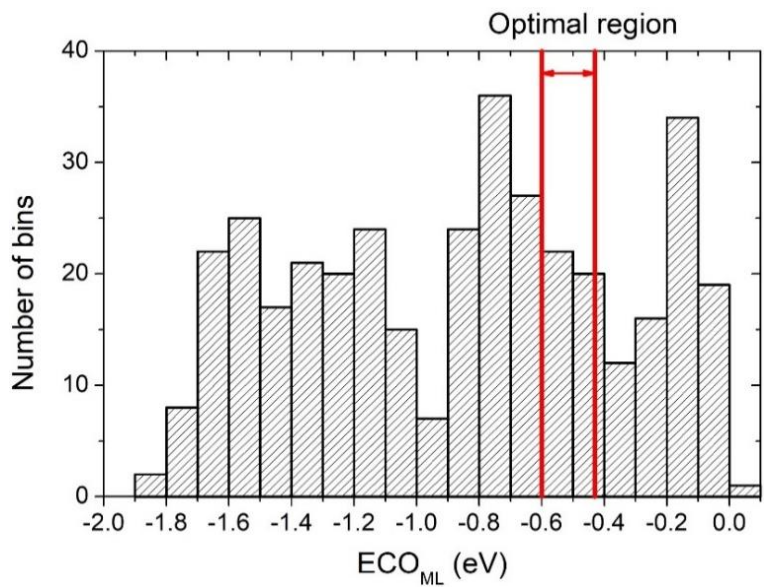

Fig. 6 Histogram of the predicted *CO binding energy ($ECO_{ML}$) using the active KRR machine using $W_d^{LMTO}$ and $\chi_P$ as descriptors after screening 372 transition metal-based alloys.

As shown in Fig.6, our actively learned chemisorption model yielded 36 candidates within the optimal target window (-0.60 ~ -0.43 eV). Among them, we chose the alloys in which the outermost surface layer is covered by Cu or Au that are nearest to the volcano tops, yielding 15 candidates for further validations (see Fig. 7). Fig. 7a shows the good agreement between $ECO_{DFT}$ and $ECO_{ML}$, indeed, all within the desired binding affinity window. In Fig. 7b, we plotted $ECOOH_{DFT}$ vs. $ECO_{DFT}$, two important reaction intermediates that determine thermodynamic limitations for the $CO_2$ reduction. Noticeably, $Cu_3Zr$@Cu and $Cu_3Y$@Cu deviate negatively from the usual scaling relation of $ECOOH_{DFT}$ and $ECO_{DFT}$, allowing *COOH formation at less negative potentials. However, $Cu_3Zr$@Cu can be ruled out since its $ECO_{DFT}$ is comparable to Cu (100), which has been observed to further protonate *CO[45]. We note that $ECO_{DFT}$ of $Cu_3Y$@Cu is 0.11 eV weaker than Cu (100), making the *CO desorption step nearly thermo-neutral (Fig. 7c). Considering the limiting potential ($U_L=-\Delta G_{MAX}/e$) as a measure of CRR activity, free energy diagram indicates that the $U_L$ of $Cu_3Y$@Cu (-0.58 V) is less negative than Au (100) (-0.95 V) by 0.37 V. Particularly, the $U_L$ of identified catalyst is comparable to other Au-based catalysts, such as Au-Cu bi-functional interfacial catalyst ($U_L$ = -0.60 V)[46] and Au NP corner site ($U_L$ = -0.60 V)[47]. The results of $Cu_3Y$@Cu should also be compared to the experimental potentials of the best performing Au-based catalysts to reach the current density of CO production of more than 5 mA/cm$^2$ in literature; -0.40 V for oxide-derived Au nanoparticles[48], -0.35 V for Au needles[49], and -0.35 V for Au nanowires[50]. All these results imply that Cu$_3$Y@Cu could be highly active, comparable to the Au catalysts, and cost-effective alternative to the Au catalysts for CO$_2$ reduction reaction.

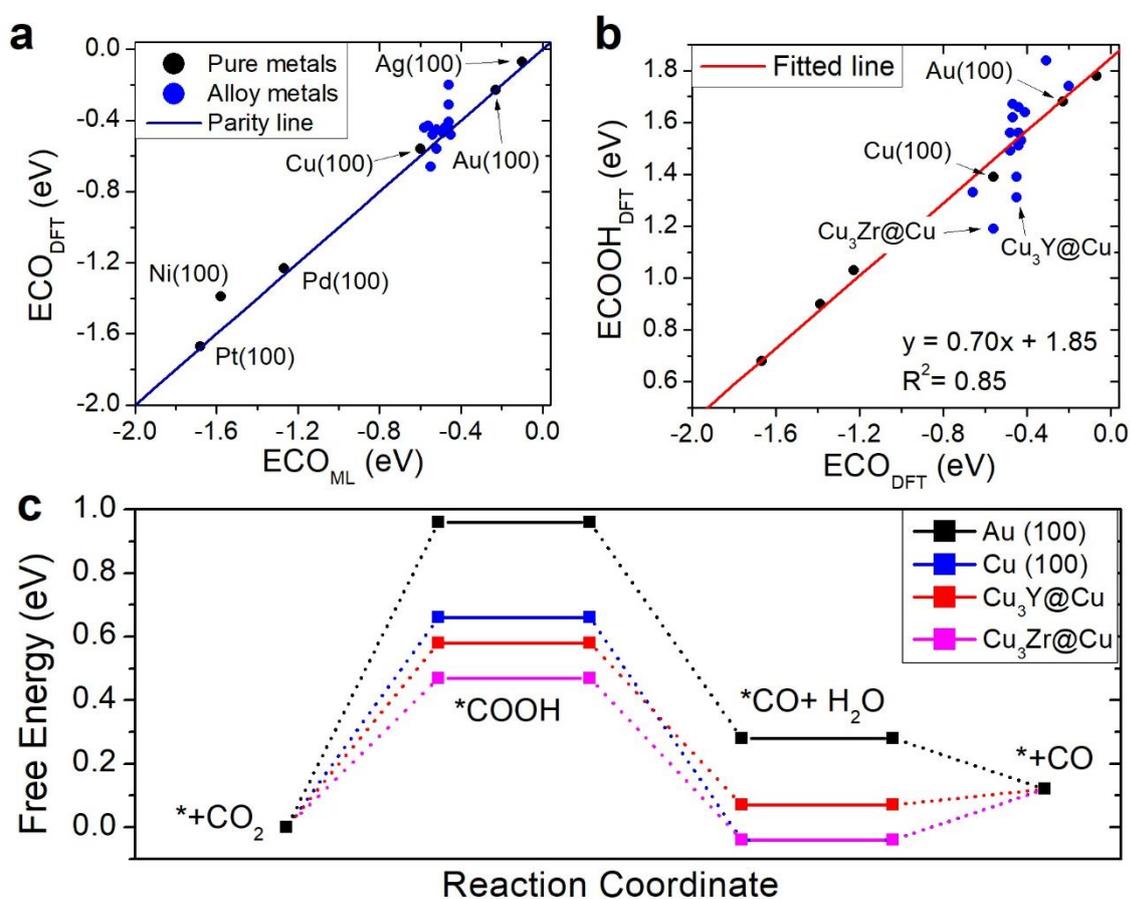

Fig. 7 (a) *CO binding energy comparison between DFT calculation ($ECO_{DFT}$) and prediction by machine learning ($ECO_{ML}$), (b) scaling relation between *CO binding energy ($ECO_{DFT}$) and *COOH binding energy ($ECOOH_{DFT}$) for data shown in (a). (c) Free energy diagram of selected catalysts. Pure Au (100) and Cu (100) surfaces are also plotted as references.

## 5. Conclusions

We presented a machine learning model that can predict the binding energy of surface adsorbates on alloys using simple non-*ab initio* input features, namely, liner muffin-tin orbital theory (LMTO)-based d-band width and a geometric mean of electronegativity. By combining the latter descriptors with the active learning algorithm, we obtained a high accuracy (RMSE = 0.05 eV for active KRR with label) to predict *CO binding energy. Use of LMTO d-band width as a learning descriptor yielded a higher prediction accuracy than the DFT-based d-band width due to the local characteristics of the $W_d^{LMTO}$. Effects of active learning were significant, lowering the RMSE for neural network 0.21 → 0.17 eV, and for KRR 0.18 → 0.05 eV. As a practical application of the constructed KRR machine, we then screened the alloy catalysts for $CO_2$ electroreduction reaction by estimating their *CO binding energies, and identified that $Cu_3Y$@Cu has the overpotential of 0.37 V lower than the pure Au (100) by deviating from the usual scaling relation. We expect that the non-*ab initio* descriptors proposed here can be easily applicable to other types of catalyst designs.

# Electronic Supplementary Information (ESI) for

# Actively learned machine with non-*ab initio* input features towards efficient $CO_2$ reduction catalyst

Juhwan Noh, Jaehoon Kim, Seoin Back, and Yousung Jung*

*Graduate School of EEWS, Korea Advanced Institute of Science and Technology (KAIST),*

*291 Daehakro, Daejeon 305-701, Korea*

*E-mail: ysjn@kaist.ac.kr*

**Brief overview of LMTO theory and d-band width calculation**

In Muffin-Tin Orbital (MTO) theory, the potential in the crystal is approximated by spherically symmetric atomic potential within a sphere of radius $r_{mt}$ (muffin-tin radius) and a constant potential region (muffin-tin potential) between the spherical atomic orbitals[1]. Based on this MTO concept, the interatomic coupling of d-states can be calculated as following expressions [1]:

$$V_{f(d,d)} = \sum_k \frac{\langle d|\Delta|k\rangle\langle k|\Delta|d\rangle}{E_d - E_k} = \eta_{f(d,d)} \frac{\hbar^2 r_d^3}{m d_{ij}^5}, \tag{S1}$$

$$\Delta = \delta V - \langle d|\delta V|d\rangle, \tag{S2}$$

$$\delta V(r) = \begin{cases} 0, & r \leq r_d \\ E_d - V_a(r), & r > r_d \end{cases}. \tag{S3}$$

In Eq. S1, $r_d$ is a spatial extent of d-orbital taken from ref.[2], $d_{ij}$ is an interatomic distance, and $f(d,d)$ is the classifier of interactions; for example, $\sigma(d,d)$ denotes the $\sigma$-interaction, and $\pi(d,d)$ denotes the $\pi$-interaction. Also, $\eta_{\sigma(d,d)}$, $\eta_{\pi(d,d)}$ and $\hbar^2/m$ are constant which can be taken from ref.[1]. Calculating Eq. S1 between different atom types can be done by changing $r_d^3$ to $r_{i,d}^{3/2} r_{j,d}^{3/2}$. $\Delta$ is a hybridization potential as shown in Eq. S2 indicating a constant shift of the potential difference function ($\delta V$), where $\delta V$ is a potential difference between the constant d-state ($E_d$) and the potential made by nucleus ($V_a(r)$) as shown in Eq. S3. Finally, using Eq. S1 with additional mathematical treatments, the d-band width ($W_d^{LMTO}$) can be obtained as following [1]:

$$W_d^{LMTO} = -\frac{8}{3} V_{\sigma(d,d)}^{1st} + \frac{32}{9} V_{\pi(d,d)}^{1st} - 3 V_{\sigma(d,d)}^{2nd} + 4 V_{\pi(d,d)}^{2nd}. \tag{S4}$$

In Eq. S4, superscripts 1st and 2nd denote the first and second nearest neighbor atoms based on interatomic distances, and we calculated Eq. S4 layer by layer as shown in Eq. S5.

$$W_d^{LMTO} = W_{d,surf.}^{LMTO} + W_{d,sub.}^{LMTO} \tag{S5}$$

Furthermore, when considering $W_{d,surf.}^{LMTO}$, the lattice parameter of surface atom was used to calculate interatomic distance, and when considering $W_{d,sub.}^{LMTO}$, the lattice parameter was obtained from Vegard's law [3] as shown in Eq. S6.

$$a_{M,X} = xa_M + (1-x)a_X, \text{ where } x = n_M/(n_M + n_X) \tag{S6}$$

In Eq. S6, $a_M$ is the lattice paramter of atom M, and $a_X$ is the lattice parameter of atom X. In addition, $n_M$ and $n_X$ are the number of M and X existing in topmost two layers. Based on these, calculations of each term in Eq. S4 are following:

Table S1 Calculation of the interatomic coupling terms for the M₃X@M alloy structure.

| M₃X@M | Surface | Sub-surface ($x = \frac{n_M}{n_M+n_X} = 0.75$) |
|---|---|---|
| $V^{1st}_{\sigma(d,d)}$ | $4 * \eta_{\sigma(d,d)} \frac{\hbar^2 r_{d,M}^{3/2} r_{d,M}^{3/2}}{ma_M^5}$ | $2 * \eta_{\pi(d,d)} \frac{\hbar^2 r_{d,M}^{\frac{3}{2}} r_{d,M}^{\frac{3}{2}}}{ma_{M,X}^5} + 2 * \eta_{\pi(d,d)} \frac{\hbar^2 r_{d,M}^{\frac{3}{2}} r_{d,X}^{\frac{3}{2}}}{ma_{M,X}^5}$ |
| $V^{2nd}_{\sigma(d,d)}$ | $4 * k_1 \eta_{\sigma(d,d)} \frac{\hbar^2 r_{d,M}^{3/2} r_{d,M}^{3/2}}{m(\sqrt{2}a_M)^5}$ | $\frac{1}{2}\left[ 4 * \eta_{\sigma(d,d)} \frac{\hbar^2 r_{d,M}^{\frac{3}{2}} r_{d,M}^{\frac{3}{2}}}{m(\sqrt{3}a_{M,X})^5} + 4 * \eta_{\sigma(d,d)} \frac{\hbar^2 r_{d,M}^{\frac{3}{2}} r_{d,X}^{\frac{3}{2}}}{m(\sqrt{3}a_{M,X})^5} \right]$ |
| $V^{1st}_{\pi(d,d)}$ | $4 * \eta_{\pi(d,d)} \frac{\hbar^2 r_{d,M}^{3/2} r_{d,M}^{3/2}}{ma_M^5}$ | $k_2 \left[ 2 * \eta_{\pi(d,d)} \frac{\hbar^2 r_{d,M}^{\frac{3}{2}} r_{d,M}^{\frac{3}{2}}}{ma_{M,X}^5} + 2 * \eta_{\pi(d,d)} \frac{\hbar^2 r_{d,M}^{\frac{3}{2}} r_{d,X}^{\frac{3}{2}}}{ma_{M,X}^5} \right]$ |
| $V^{2nd}_{\pi(d,d)}$ | $4 * k_1 \eta_{\pi(d,d)} \frac{\hbar^2 r_{d,M}^{3/2} r_{d,M}^{3/2}}{m(\sqrt{2}a_M)^5}$ | $\frac{k_2}{2}\left[ 4 * \eta_{\pi(d,d)} \frac{\hbar^2 r_{d,M}^{\frac{3}{2}} r_{d,M}^{\frac{3}{2}}}{m(\sqrt{3}a_{M,X})^5} + 4 * \eta_{\pi(d,d)} \frac{\hbar^2 r_{d,M}^{\frac{3}{2}} r_{d,X}^{\frac{3}{2}}}{m(\sqrt{3}a_{M,X})^5} \right]$ |

Table S2 Calculation of the interatomic coupling terms for the M-X@M and X@M alloy structures.

| M-X@M and X@M | Surface | Sub-surface ($x = \frac{n_M}{n_M+n_X} = 0.5$) |
|---|---|---|
| $V^{1st}_{\sigma(d,d)}$ | $4 * \eta_{\sigma(d,d)} \frac{\hbar^2 r_{d,M}^{3/2} r_{d,M}^{3/2}}{ma_M^5}$ | $4 * \eta_{\sigma(d,d)} \frac{\hbar^2 r_{d,M}^{\frac{3}{2}} r_{d,X}^{\frac{3}{2}}}{ma_{M,X}^5}$ |
| $V^{2nd}_{\sigma(d,d)}$ | $4 * k_1 \eta_{\sigma(d,d)} \frac{\hbar^2 r_{d,M}^{3/2} r_{d,M}^{3/2}}{m(\sqrt{2}a_M)^5}$ | $\frac{1}{2}\left[ 8 * \eta_{\sigma(d,d)} \frac{\hbar^2 r_{d,M}^{\frac{3}{2}} r_{d,X}^{\frac{3}{2}}}{m(\sqrt{3}a_{M,X})^5} \right]$ |
| $V^{1st}_{\pi(d,d)}$ | $4 * \eta_{\pi(d,d)} \frac{\hbar^2 r_{d,M}^{3/2} r_{d,M}^{3/2}}{ma_M^5}$ | $k_2 \left[ 4 * \eta_{\pi(d,d)} \frac{\hbar^2 r_{d,M}^{\frac{3}{2}} r_{d,X}^{\frac{3}{2}}}{ma_{M,X}^5} \right]$ |
| $V^{2nd}_{\pi(d,d)}$ | $4 * k_1 \eta_{\pi(d,d)} \frac{\hbar^2 r_{d,M}^{3/2} r_{d,M}^{3/2}}{m(\sqrt{2}a_M)^5}$ | $\frac{k_2}{2}\left[ 8 * \eta_{\pi(d,d)} \frac{\hbar^2 r_{d,M}^{\frac{3}{2}} r_{d,X}^{\frac{3}{2}}}{m(\sqrt{3}a_{M,X})^5} \right]$ |

In the tables, $r_{d,M}$ and $r_{d,X}$ are the spatial extents of d-state for atom M and X, respectively, and for both $k_1$ and $k_2$ we explicitly used $k_1 = 16$ and $k_2 = -1.85$. In addition, all multiplied integers denote the number of coordinated atoms.

Table S3 Descriptors used for machine learning models

| Alloy name | This work | | | Taken from the ref.[4] | | |
|---|---|---|---|---|---|---|
| | $\chi_M$ | $\chi_P$ | $W_d^{LMTO}$ (eV) | $W_d^{cal}$ (eV) | $d_c$ (eV) | $ECO_{cal}$ (eV) |
| Ag3Ag@Ag | 4.44 | 1.93 | 31.37 | 1.10 | -3.92 | -0.09 |
| Ag3Au@Ag | 4.74 | 2.07 | 31.19 | 1.13 | -3.86 | -0.11 |
| Ag3Co@Ag | 4.40 | 1.92 | 30.56 | 1.19 | -3.90 | -0.11 |
| Ag3Cr@Ag | 4.25 | 1.86 | 30.58 | 1.13 | -4.03 | -0.12 |
| Ag3Cu@Ag | 4.45 | 1.92 | 31.43 | 1.16 | -3.99 | -0.08 |
| Ag3Ir@Ag | 4.65 | 1.99 | 30.94 | 1.22 | -3.66 | -0.14 |
| Ag3Mn@Ag | 4.25 | 1.83 | 32.42 | 1.15 | -4.08 | -0.12 |
| Ag3Nb@Ag | 4.28 | 1.84 | 30.14 | 1.14 | -3.89 | -0.07 |
| Ag3Ni@Ag | 4.43 | 1.92 | 31.33 | 1.20 | -3.85 | -0.09 |
| Ag3Os@Ag | 4.55 | 1.99 | 29.94 | 1.24 | -3.74 | -0.10 |
| Ag3Pd@Ag | 4.44 | 1.99 | 31.19 | 1.13 | -3.69 | -0.08 |
| Ag3Pt@Ag | 4.70 | 2.01 | 31.05 | 1.18 | -3.68 | -0.12 |
| Ag3Ru@Ag | 4.38 | 1.99 | 30.09 | 1.18 | -3.65 | -0.11 |
| Ag3Sc@Ag | 4.15 | 1.77 | 30.24 | 1.06 | -4.02 | -0.05 |
| Ag3Ta@Ag | 4.36 | 1.81 | 30.02 | 1.21 | -3.98 | -0.02 |
| Ag3Ti@Ag | 4.17 | 1.82 | 30.26 | 1.09 | -4.01 | -0.10 |
| Ag3V@Ag | 4.22 | 1.85 | 30.53 | 1.13 | -3.98 | -0.12 |
| Ag3W@Ag | 4.43 | 2.03 | 30.03 | 1.24 | -3.89 | -0.03 |
| Au3Ag@Au | 5.40 | 2.37 | 46.62 | 1.37 | -3.17 | -0.21 |
| Au3Au@Au | 5.77 | 2.54 | 46.40 | 1.41 | -3.18 | -0.26 |
| Au3Cu@Au | 5.42 | 2.36 | 46.67 | 1.48 | -3.34 | -0.17 |
| Au3Pd@Au | 5.41 | 2.45 | 46.39 | 1.43 | -2.99 | -0.30 |
| Au3Pt@Au | 5.72 | 2.47 | 46.22 | 1.47 | -3.02 | -0.36 |
| Cu3Ag@Cu | 4.48 | 1.91 | 24.88 | 0.96 | -2.09 | -0.70 |
| Cu3Au@Cu | 4.78 | 2.04 | 24.70 | 1.02 | -2.11 | -0.72 |
| Cu3Co@Cu | 4.43 | 1.89 | 24.14 | 1.07 | -2.19 | -0.63 |
| Cu3Cr@Cu | 4.28 | 1.84 | 24.12 | 1.01 | -2.36 | -0.66 |
| Cu3Cu@Cu | 4.49 | 1.90 | 24.98 | 1.06 | -2.26 | -0.63 |
| Cu3Ir@Cu | 4.69 | 1.97 | 24.45 | 1.10 | -2.09 | -0.72 |
| Cu3La@Cu | 4.17 | 1.66 | 23.41 | 0.86 | -2.16 | -0.60 |
| Cu3Mn@Cu | 4.28 | 1.81 | 25.83 | 1.01 | -2.38 | -0.63 |
| Cu3Mo@Cu | 4.34 | 1.96 | 23.68 | 1.03 | -2.27 | -0.48 |
| Cu3Nb@Cu | 4.32 | 1.82 | 23.64 | 0.99 | -2.36 | -0.51 |
| Cu3Ni@Cu | 4.47 | 1.90 | 24.88 | 1.07 | -2.15 | -0.66 |
| Cu3Os@Cu | 4.59 | 1.97 | 23.44 | 1.13 | -2.23 | -0.65 |
| Cu3Pd@Cu | 4.48 | 1.97 | 24.71 | 1.01 | -1.94 | -0.70 |
| Cu3Pt@Cu | 4.74 | 1.99 | 24.56 | 1.07 | -1.97 | -0.74 |
| Cu3Re@Cu | 4.37 | 1.90 | 23.28 | 1.13 | -2.31 | -0.49 |
| Cu3Rh@Cu | 4.44 | 1.99 | 24.58 | 1.04 | -1.98 | -0.72 |
| Cu3Ru@Cu | 4.42 | 1.97 | 23.61 | 1.04 | -2.04 | -0.67 |
| Cu3Sc@Cu | 4.18 | 1.75 | 23.73 | 0.91 | -2.38 | -0.51 |
| Cu3Ta@Cu | 4.39 | 1.79 | 23.51 | 1.03 | -2.47 | -0.49 |
| Cu3Ti@Cu | 4.21 | 1.80 | 23.77 | 0.97 | -2.45 | -0.52 |
| Cu3V@Cu | 4.26 | 1.83 | 24.06 | 1.00 | -2.34 | -0.65 |
| Cu3W@Cu | 4.47 | 2.01 | 23.52 | 1.08 | -2.38 | -0.46 |
| Cu3Y@Cu | 4.15 | 1.70 | 23.31 | 0.90 | -2.31 | -0.49 |
| Cu3Zr@Cu | 4.23 | 1.74 | 23.26 | 0.93 | -2.39 | -0.56 |
| Ni3Ag@Ni | 4.41 | 1.91 | 33.75 | 1.22 | -1.10 | -1.62 |
| Ni3Au@Ni | 4.71 | 2.05 | 33.53 | 1.27 | -1.12 | -1.68 |
| Ni3Co@Ni | 4.37 | 1.90 | 32.79 | 1.36 | -1.42 | -1.47 |
| Ni3Cr@Ni | 4.22 | 1.84 | 32.79 | 1.33 | -1.31 | -1.53 |
| Ni3Cu@Ni | 4.42 | 1.91 | 33.85 | 1.29 | -1.19 | -1.52 |
| Ni3Ir@Ni | 4.62 | 1.98 | 33.22 | 1.37 | -1.30 | -1.58 |
| Ni3La@Ni | 4.10 | 1.66 | 31.98 | 1.07 | -0.75 | -1.47 |
| Ni3Mo@Ni | 4.27 | 1.97 | 32.27 | 1.37 | -1.52 | -1.36 |
| Ni3Nb@Ni | 4.25 | 1.83 | 32.23 | 1.34 | -1.49 | -1.35 |
| Ni3Ni@Ni | 4.40 | 1.91 | 33.73 | 1.33 | -1.32 | -1.52 |
| Ni3Os@Ni | 4.52 | 1.98 | 31.96 | 1.43 | -1.45 | -1.44 |
| Ni3Pd@Ni | 4.41 | 1.98 | 33.53 | 1.27 | -1.20 | -1.61 |
| Ni3Pt@Ni | 4.67 | 2.00 | 33.36 | 1.33 | -1.25 | -1.65 |
| Ni3Re@Ni | 4.30 | 1.91 | 31.77 | 1.46 | -1.59 | -1.34 |
| Ni3Rh@Ni | 4.37 | 2.00 | 33.37 | 1.32 | -1.30 | -1.56 |
| Ni3Ru@Ni | 4.35 | 1.98 | 32.16 | 1.32 | -1.33 | -1.52 |
| Ni3Sc@Ni | 4.12 | 1.75 | 32.34 | 1.18 | -0.94 | -1.47 |
| Ni3Ta@Ni | 4.33 | 1.80 | 32.07 | 1.40 | -1.54 | -1.33 |
| Ni3Ti@Ni | 4.14 | 1.81 | 32.37 | 1.30 | -1.20 | -1.52 |
| Ni3V@Ni | 4.20 | 1.84 | 32.72 | 1.37 | -1.52 | -1.40 |
| Ni3W@Ni | 4.40 | 2.01 | 32.08 | 1.43 | -1.60 | -1.34 |
| Ni3Y@Ni | 4.09 | 1.71 | 31.85 | 1.10 | -0.79 | -1.49 |
| Ni3Zr@Ni | 4.16 | 1.74 | 31.77 | 1.23 | -1.04 | -1.54 |
| Pd3Ag@Pd | 4.45 | 2.13 | 47.73 | 1.29 | -1.53 | -1.24 |
| Pd3Au@Pd | 4.75 | 2.28 | 47.50 | 1.35 | -1.54 | -1.34 |
| Pd3Co@Pd | 4.40 | 2.12 | 46.55 | 1.48 | -1.95 | -1.07 |
| Pd3Cr@Pd | 4.26 | 2.05 | 46.60 | 1.43 | -1.87 | -0.99 |
| Pd3Cu@Pd | 4.46 | 2.12 | 47.79 | 1.40 | -1.72 | -1.15 |
| Pd3Ir@Pd | 4.66 | 2.20 | 47.15 | 1.50 | -1.93 | -1.20 |
| Pd3La@Pd | 4.14 | 1.85 | 45.84 | 1.25 | -1.48 | -0.80 |
| Pd3Mn@Pd | 4.26 | 2.02 | 49.19 | 1.41 | -1.75 | -1.00 |
| Pd3Mo@Pd | 4.31 | 2.19 | 46.10 | 1.56 | -2.36 | -0.94 |
| Pd3Nb@Pd | 4.29 | 2.03 | 46.07 | 1.58 | -2.59 | -0.80 |
| Pd3Ni@Pd | 4.44 | 2.12 | 47.65 | 1.43 | -1.82 | -1.17 |
| Pd3Os@Pd | 4.56 | 2.20 | 45.74 | 1.59 | -2.20 | -1.08 |
| Pd3Pd@Pd | 4.45 | 2.20 | 47.48 | 1.36 | -1.67 | -1.28 |
| Pd3Pt@Pd | 4.71 | 2.22 | 47.30 | 1.42 | -1.72 | -1.31 |
| Pd3Re@Pd | 4.34 | 2.12 | 45.55 | 1.65 | -2.45 | -0.93 |
| Pd3Rh@Pd | 4.41 | 2.22 | 47.31 | 1.41 | -1.82 | -1.24 |
| Pd3Sc@Pd | 4.15 | 1.95 | 46.19 | 1.39 | -1.72 | -0.82 |
| Pd3Ta@Pd | 4.36 | 2.00 | 45.91 | 1.67 | -2.76 | -0.76 |
| Pd3Ti@Pd | 4.18 | 2.01 | 46.18 | 1.51 | -2.27 | -0.94 |
| Pd3V@Pd | 4.23 | 2.04 | 46.55 | 1.51 | -2.19 | -0.95 |
| Pd3W@Pd | 4.44 | 2.24 | 45.91 | 1.68 | -2.67 | -0.83 |
| Pd3Y@Pd | 4.12 | 1.90 | 45.70 | 1.31 | -1.51 | -0.82 |
| Pd3Zr@Pd | 4.20 | 1.94 | 45.59 | 1.51 | -2.17 | -0.89 |
| Pt3Ag@Pt | 5.26 | 2.19 | 62.37 | 1.74 | -1.79 | -1.58 |
| Pt3Au@Pt | 5.62 | 2.34 | 62.11 | 1.77 | -1.80 | -1.72 |
| Pt3Co@Pt | 5.21 | 2.17 | 60.90 | 1.95 | -2.25 | -1.44 |
| Pt3Cr@Pt | 5.04 | 2.11 | 61.00 | 1.92 | -2.23 | -1.39 |
| Pt3Cu@Pt | 5.28 | 2.18 | 62.41 | 1.88 | -2.03 | -1.44 |
| Pt3Ir@Pt | 5.51 | 2.26 | 61.70 | 1.95 | -2.18 | -1.62 |
| Pt3La@Pt | 4.90 | 1.90 | 60.21 | 1.54 | -1.56 | -1.25 |
| Pt3Mo@Pt | 5.10 | 2.25 | 60.46 | 2.00 | -2.55 | -1.32 |
| Pt3Nb@Pt | 5.07 | 2.09 | 60.44 | 2.01 | -2.59 | -1.20 |
| Pt3Ni@Pt | 5.25 | 2.18 | 62.24 | 1.91 | -2.16 | -1.51 |
| Pt3Os@Pt | 5.39 | 2.26 | 60.02 | 2.03 | -2.40 | -1.51 |
| Pt3Pd@Pt | 5.27 | 2.26 | 62.08 | 1.81 | -1.97 | -1.65 |
| Pt3Pt@Pt | 5.57 | 2.28 | 61.88 | 1.87 | -2.03 | -1.70 |
| Pt3Re@Pt | 5.13 | 2.18 | 59.80 | 2.10 | -2.58 | -1.38 |
| Pt3Rh@Pt | 5.22 | 2.28 | 61.88 | 1.87 | -2.12 | -1.61 |
| Pt3Ru@Pt | 5.19 | 2.26 | 60.23 | 1.91 | -2.22 | -1.57 |
| Pt3Sc@Pt | 4.92 | 2.00 | 60.58 | 1.83 | -2.00 | -1.23 |
| Pt3Ta@Pt | 5.16 | 2.05 | 60.26 | 2.11 | -2.73 | -1.15 |
| Pt3Ti@Pt | 4.94 | 2.07 | 60.54 | 1.96 | -2.45 | -1.31 |
| Pt3V@Pt | 5.01 | 2.10 | 60.95 | 1.98 | -2.47 | -1.37 |
| Pt3W@Pt | 5.26 | 2.30 | 60.25 | 2.13 | -2.72 | -1.18 |
| Pt3Y@Pt | 4.88 | 1.95 | 60.05 | 1.68 | -1.71 | -1.24 |
| Pt3Zr@Pt | 4.97 | 1.99 | 59.89 | 1.89 | -2.19 | -1.33 |
| Ag-Ag@Ag | 4.44 | 1.93 | 31.37 | 1.10 | -3.92 | -0.09 |
| Ag-Au@Ag | 5.06 | 2.21 | 31.01 | 1.17 | -3.86 | -0.12 |
| Ag-Cu@Ag | 4.46 | 1.91 | 31.56 | 1.18 | -4.01 | -0.06 |
| Ag-La@Ag | 3.85 | 1.46 | 28.21 | 1.06 | -4.15 | -0.12 |
| Ag-Nb@Ag | 4.12 | 1.76 | 28.31 | 1.26 | -4.33 | -0.24 |
| Ag-Ni@Ag | 4.42 | 1.92 | 31.33 | 1.20 | -3.73 | -0.14 |
| Ag-Pd@Ag | 4.44 | 2.06 | 30.99 | 1.13 | -3.50 | -0.15 |
| Ag-Pt@Ag | 4.97 | 2.10 | 30.70 | 1.23 | -3.55 | -0.23 |
| Ag-Sc@Ag | 3.87 | 1.62 | 28.55 | 1.17 | -4.49 | -0.06 |
| Ag-Ta@Ag | 4.27 | 1.70 | 28.01 | 1.35 | -4.41 | -0.27 |
| Ag-Ti@Ag | 3.92 | 1.72 | 28.36 | 1.16 | -4.41 | -0.15 |
| Ag-Y@Ag | 3.81 | 1.53 | 27.89 | 1.14 | -4.36 | -0.11 |
| Ag-Zr@Ag | 3.96 | 1.60 | 27.31 | 1.21 | -4.42 | -0.13 |
| Au-Ag@Au | 5.06 | 2.21 | 46.84 | 1.32 | -3.19 | -0.18 |
| Au-Au@Au | 5.77 | 2.54 | 46.40 | 1.41 | -3.18 | -0.26 |
| Au-Cu@Au | 5.09 | 2.20 | 47.07 | 1.45 | -3.40 | -0.11 |
| Au-La@Au | 4.38 | 1.67 | 43.00 | 1.36 | -3.71 | -0.11 |
| Au-Pd@Au | 5.07 | 2.36 | 46.38 | 1.39 | -2.89 | -0.39 |
| Au-Pt@Au | 5.67 | 2.41 | 46.03 | 1.49 | -3.01 | -0.51 |
| Au-Ru@Au | 4.93 | 2.36 | 42.34 | 1.49 | -3.35 | -0.47 |
| Au-Y@Au | 4.34 | 1.76 | 42.61 | 1.47 | -4.01 | -0.10 |
| Cu-Ag@Cu | 4.46 | 1.91 | 24.84 | 0.98 | -2.13 | -0.64 |
| Cu-Au@Cu | 5.09 | 2.20 | 24.52 | 1.04 | -2.08 | -0.67 |
| Cu-Co@Cu | 4.38 | 1.89 | 22.67 | 1.08 | -2.34 | -0.76 |
| Cu-Cr@Cu | 4.09 | 1.78 | 22.82 | 1.10 | -2.59 | -1.17 |
| Cu-Cu@Cu | 4.49 | 1.90 | 24.98 | 1.06 | -2.26 | -0.63 |
| Cu-Ir@Cu | 4.90 | 2.04 | 24.00 | 1.16 | -2.31 | -0.78 |
| Cu-La@Cu | 3.87 | 1.45 | 21.99 | 0.93 | -2.41 | -0.69 |
| Cu-Mo@Cu | 4.20 | 2.03 | 21.94 | 1.11 | -2.52 | -1.18 |
| Cu-Nb@Cu | 4.15 | 1.74 | 21.99 | 1.08 | -2.71 | -1.05 |
| Cu-Ni@Cu | 4.44 | 1.90 | 24.77 | 1.07 | -2.14 | -0.72 |
| Cu-Os@Cu | 4.69 | 2.04 | 20.85 | 1.19 | -2.52 | -0.90 |
| Cu-Pd@Cu | 4.47 | 2.04 | 24.49 | 1.03 | -1.99 | -0.75 |
| Cu-Pt@Cu | 5.00 | 2.08 | 24.24 | 1.12 | -2.09 | -0.79 |
| Cu-Re@Cu | 4.25 | 1.90 | 20.43 | 1.21 | -2.64 | -1.12 |
| Cu-Rh@Cu | 4.39 | 2.08 | 24.23 | 1.07 | -2.14 | -0.81 |
| Cu-Ru@Cu | 4.35 | 2.04 | 21.26 | 1.08 | -2.31 | -0.87 |
| Cu-Sc@Cu | 3.90 | 1.61 | 22.22 | 0.99 | -2.72 | -0.63 |
| Cu-Ta@Cu | 4.30 | 1.69 | 21.72 | 1.14 | -2.83 | -1.01 |
| Cu-Ti@Cu | 3.94 | 1.71 | 21.97 | 1.02 | -2.83 | -0.69 |
| Cu-V@Cu | 4.04 | 1.76 | 22.75 | 1.07 | -2.74 | -0.81 |
| Cu-W@Cu | 4.45 | 2.12 | 21.58 | 1.19 | -2.63 | -1.18 |
| Cu-Y@Cu | 3.83 | 1.52 | 21.69 | 0.96 | -2.56 | -0.73 |
| Cu-Zr@Cu | 3.98 | 1.59 | 21.06 | 1.02 | -2.74 | -0.84 |
| Ni-Ag@Ni | 4.42 | 1.92 | 33.81 | 1.18 | -1.08 | -1.61 |
| Ni-Au@Ni | 5.04 | 2.20 | 33.45 | 1.24 | -1.08 | -1.71 |
| Ni-Co@Ni | 4.33 | 1.89 | 31.24 | 1.37 | -1.44 | -1.45 |
| Ni-Cr@Ni | 4.05 | 1.78 | 31.44 | 1.38 | -1.53 | -1.39 |
| Ni-Cu@Ni | 4.44 | 1.90 | 33.97 | 1.26 | -1.11 | -1.54 |
| Ni-Ir@Ni | 4.85 | 2.05 | 32.84 | 1.45 | -1.36 | -1.80 |
| Ni-Mo@Ni | 4.15 | 2.03 | 30.41 | 1.40 | -1.35 | -1.65 |
| Ni-Nb@Ni | 4.11 | 1.75 | 30.49 | 1.37 | -1.27 | -1.47 |
| Ni-Ni@Ni | 4.40 | 1.91 | 33.73 | 1.33 | -1.32 | -1.52 |
| Ni-Os@Ni | 4.64 | 2.05 | 29.12 | 1.53 | -1.50 | -1.87 |
| Ni-Pd@Ni | 4.42 | 2.05 | 33.41 | 1.27 | -1.22 | -1.68 |
| Ni-Pt@Ni | 4.95 | 2.09 | 33.12 | 1.36 | -1.28 | -1.74 |
| Ni-Re@Ni | 4.21 | 1.90 | 28.63 | 1.54 | -1.50 | -1.75 |
| Ni-Rh@Ni | 4.35 | 2.09 | 33.10 | 1.34 | -1.30 | -1.73 |
| Ni-Ru@Ni | 4.30 | 2.05 | 29.60 | 1.36 | -1.35 | -1.87 |
| Ni-Sc@Ni | 3.86 | 1.61 | 30.75 | 1.28 | -1.22 | -1.27 |
| Ni-Ta@Ni | 4.25 | 1.69 | 30.17 | 1.43 | -1.35 | -1.44 |
| Ni-Ti@Ni | 3.90 | 1.72 | 30.44 | 1.33 | -1.36 | -1.30 |
| Ni-V@Ni | 4.00 | 1.76 | 31.36 | 1.36 | -1.43 | -1.36 |
| Ni-W@Ni | 4.41 | 2.12 | 30.00 | 1.50 | -1.43 | -1.62 |
| Ni-Y@Ni | 3.79 | 1.53 | 30.15 | 1.20 | -1.10 | -1.23 |
| Ni-Zr@Ni | 3.94 | 1.59 | 29.40 | 1.27 | -1.19 | -1.33 |
| Pd-Ag@Pd | 4.44 | 2.06 | 47.95 | 1.28 | -1.40 | -1.16 |
| Pd-Au@Pd | 5.07 | 2.36 | 47.51 | 1.39 | -1.51 | -1.31 |
| Pd-Co@Pd | 4.36 | 2.03 | 45.16 | 1.47 | -2.06 | -0.90 |
| Pd-Cu@Pd | 4.47 | 2.04 | 48.17 | 1.38 | -1.64 | -1.12 |
| Pd-Ir@Pd | 4.87 | 2.20 | 46.81 | 1.63 | -2.23 | -1.23 |
| Pd-La@Pd | 3.85 | 1.56 | 44.04 | 1.32 | -1.96 | -0.50 |
| Pd-Mn@Pd | 4.07 | 1.85 | 49.67 | 1.49 | -2.03 | -0.68 |
| Pd-Mo@Pd | 4.18 | 2.18 | 44.06 | 1.61 | -2.46 | -0.79 |
| Pd-Nb@Pd | 4.13 | 1.88 | 44.11 | 1.62 | -2.47 | -0.72 |
| Pd-Ni@Pd | 4.42 | 2.05 | 47.89 | 1.43 | -1.90 | -1.05 |
| Pd-Os@Pd | 4.67 | 2.20 | 42.68 | 1.71 | -2.50 | -1.05 |
| Pd-Pd@Pd | 4.45 | 2.20 | 47.48 | 1.36 | -1.67 | -1.28 |
| Pd-Pt@Pd | 4.98 | 2.24 | 47.13 | 1.49 | -1.82 | -1.33 |
| Pd-Rh@Pd | 4.37 | 2.24 | 47.12 | 1.47 | -1.99 | -1.18 |
| Pd-Ru@Pd | 4.33 | 2.20 | 43.24 | 1.51 | -2.16 | -1.09 |
| Pd-Sc@Pd | 3.88 | 1.73 | 44.42 | 1.50 | -2.39 | -0.48 |
| Pd-Ta@Pd | 4.28 | 1.82 | 43.74 | 1.71 | -2.59 | -0.74 |
| Pd-Ti@Pd | 3.92 | 1.84 | 44.14 | 1.51 | -2.56 | -0.53 |
| Pd-V@Pd | 4.02 | 1.89 | 45.18 | 1.54 | -2.53 | -0.61 |
| Pd-W@Pd | 4.43 | 2.28 | 43.58 | 1.75 | -2.67 | -0.80 |
| Pd-Y@Pd | 3.81 | 1.64 | 43.64 | 1.41 | -2.17 | -0.55 |
| Pd-Zr@Pd | 3.96 | 1.71 | 42.86 | 1.55 | -2.32 | -0.62 |
| Pt-Ag@Pt | 4.97 | 2.10 | 62.82 | 1.62 | -1.56 | -1.52 |
| Pt-Au@Pt | 5.67 | 2.41 | 62.32 | 1.74 | -1.63 | -1.67 |
| Pt-Co@Pt | 4.88 | 2.07 | 59.65 | 1.81 | -2.12 | -1.18 |
| Pt-Cu@Pt | 5.00 | 2.08 | 63.07 | 1.74 | -1.76 | -1.44 |
| Pt-Ir@Pt | 5.45 | 2.24 | 61.51 | 2.01 | -2.41 | -1.52 |
| Pt-La@Pt | 4.31 | 1.58 | 58.36 | 1.69 | -2.05 | -0.72 |
| Pt-Mo@Pt | 4.67 | 2.22 | 58.39 | 2.01 | -2.59 | -1.09 |
| Pt-Nb@Pt | 4.62 | 1.91 | 58.45 | 2.02 | -2.60 | -1.06 |
| Pt-Ni@Pt | 4.95 | 2.09 | 62.75 | 1.80 | -2.05 | -1.36 |
| Pt-Os@Pt | 5.22 | 2.24 | 56.83 | 2.10 | -2.65 | -1.40 |
| Pt-Pd@Pt | 4.98 | 2.24 | 62.28 | 1.71 | -1.86 | -1.64 |
| Pt-Pt@Pt | 5.57 | 2.28 | 61.88 | 1.87 | -2.03 | -1.70 |
| Pt-Re@Pt | 4.73 | 2.08 | 56.19 | 2.14 | -2.73 | -1.20 |
| Pt-Rh@Pt | 4.89 | 2.28 | 61.87 | 1.84 | -2.19 | -1.49 |
| Pt-Ru@Pt | 4.84 | 2.24 | 57.47 | 1.88 | -2.32 | -1.43 |
| Pt-Sc@Pt | 4.34 | 1.76 | 58.80 | 1.90 | -2.46 | -0.68 |
| Pt-Ta@Pt | 4.78 | 1.85 | 58.02 | 2.15 | -2.74 | -1.04 |
| Pt-Ti@Pt | 4.39 | 1.87 | 58.49 | 1.90 | -2.70 | -0.77 |
| Pt-V@Pt | 4.50 | 1.93 | 59.67 | 1.93 | -2.65 | -0.88 |
| Pt-W@Pt | 4.96 | 2.32 | 57.85 | 2.16 | -2.76 | -1.08 |
| Pt-Y@Pt | 4.27 | 1.67 | 57.90 | 1.80 | -2.33 | -0.79 |
| Pt-Zr@Pt | 4.43 | 1.74 | 57.02 | 1.95 | -2.51 | -0.89 |
| Ag@Ag | 4.44 | 1.93 | 31.37 | 1.10 | -3.92 | -0.09 |
| Ag@Au | 5.06 | 2.21 | 46.84 | 1.34 | -3.21 | -0.15 |
| Ag@Cu | 4.46 | 1.91 | 24.84 | 0.88 | -1.90 | -0.74 |
| Ag@Ni | 4.42 | 1.92 | 33.81 | 1.05 | -0.77 | -1.73 |
| Ag@Pd | 4.44 | 2.06 | 47.95 | 1.23 | -1.31 | -1.12 |
| Ag@Pt | 4.97 | 2.10 | 62.82 | 1.52 | -1.38 | -1.59 |
| Au@Ag | 5.06 | 2.21 | 31.01 | 1.16 | -3.87 | -0.15 |
| Au@Au | 5.77 | 2.54 | 46.40 | 1.41 | -3.18 | -0.26 |
| Au@Cu | 5.09 | 2.20 | 24.52 | 1.02 | -1.98 | -0.86 |
| Au@Ni | 5.04 | 2.20 | 33.45 | 1.19 | -0.87 | -1.80 |
| Au@Pd | 5.07 | 2.36 | 47.51 | 1.35 | -1.35 | -1.29 |
| Au@Pt | 5.67 | 2.41 | 62.32 | 1.64 | -1.46 | -1.84 |
| Cu@Au | 5.09 | 2.20 | 47.07 | 1.90 | -3.88 | -0.66 |
| Cu@Cu | 4.49 | 1.90 | 24.98 | 1.05 | -2.25 | -0.63 |
| Cu@Ni | 4.44 | 1.90 | 33.97 | 1.22 | -1.01 | -1.60 |
| Cu@Pd | 4.47 | 2.04 | 48.17 | 1.62 | -2.05 | -0.84 |
| Cu@Pt | 5.00 | 2.08 | 63.07 | 2.05 | -2.33 | -1.11 |
| Ir@Ag | 4.87 | 2.06 | 30.45 | 1.36 | -3.81 | -0.25 |
| Ir@Au | 5.55 | 2.36 | 45.72 | 1.72 | -3.41 | -0.37 |
| Ir@Cu | 4.90 | 2.04 | 24.00 | 1.13 | -2.10 | -0.91 |
| Ir@Ni | 4.85 | 2.05 | 32.84 | 1.36 | -1.21 | -1.59 |
| Ir@Pd | 4.87 | 2.20 | 46.81 | 1.60 | -2.15 | -1.18 |
| Ir@Pt | 5.45 | 2.24 | 61.51 | 2.04 | -2.37 | -1.49 |
| Ni@Cu | 4.44 | 1.90 | 24.77 | 1.11 | -2.21 | -0.69 |
| Ni@Ni | 4.40 | 1.91 | 33.73 | 1.33 | -1.32 | -1.52 |
| Pd@Ag | 4.44 | 2.06 | 30.99 | 1.22 | -3.64 | -0.14 |
| Pd@Au | 5.07 | 2.36 | 46.38 | 1.52 | -3.07 | -0.27 |
| Pd@Cu | 4.47 | 2.04 | 24.49 | 0.97 | -1.66 | -0.86 |
| Pd@Ni | 4.42 | 2.05 | 33.41 | 1.22 | -1.13 | -1.64 |
| Pd@Pd | 4.45 | 2.20 | 47.48 | 1.36 | -1.67 | -1.28 |
| Pd@Pt | 4.98 | 2.24 | 62.28 | 1.74 | -1.90 | -1.66 |
| Pt@Ag | 4.97 | 2.10 | 30.70 | 1.28 | -3.69 | -0.24 |
| Pt@Au | 5.67 | 2.41 | 46.03 | 1.59 | -3.17 | -0.40 |
| Pt@Cu | 5.00 | 2.08 | 24.24 | 1.09 | -1.85 | -0.91 |
| Pt@Ni | 4.95 | 2.09 | 33.12 | 1.34 | -1.24 | -1.63 |
| Pt@Pd | 4.98 | 2.24 | 47.13 | 1.48 | -1.82 | -1.31 |
| Pt@Pt | 5.57 | 2.28 | 61.88 | 1.87 | -2.03 | -1.70 |